\documentclass{aa}
\usepackage{graphics}

\begin{document}

   \thesaurus{03
               (13.09.3);
               (13.09.4);
               (11.09.1 M\,82);
               (11.09.1 NGC\,253);
               (11.09.1 Circinus);
               (11.09.1 NGC\,1068)} 
   \title{ISO-SWS spectra of galaxies: continuum and features
          \thanks{Based on observations with ISO, an ESA project with
                  instruments funded by ESA member states (especially
                  the PI countries: France, Germany, the Netherlands and
                  the United Kingdom) and with participation of ISAS and
                  NASA.} }

   \author{E. Sturm \inst{1,}\inst{2}
          \and{D. Lutz \inst{1}}
          \and{D. Tran \inst{1}}
          \and{H. Feuchtgruber \inst{1}}
          \and{R. Genzel \inst{1}}
          \and{D. Kunze \inst{1}}
          \and{A.F.M. Moorwood \inst{3}}
          \and{M.D. Thornley \inst{4}}}

   \offprints{sturm@mpe.mpg.de}

   \institute{Max-Planck-Institut f\"ur extraterrestrische Physik,
                 Postfach 1603, D-85740 Garching, Germany
              \and Infrared Processing and Analysis Center, 
                 MS 100-22, Pasadena, CA 91125, USA
              \and European Southern Observatory,
                 Karl-Schwarzschild-Str. 2, D-85748 Garching, Germany 
              \and National Radio Astronomy Observatory,
                 520 Edgemont Road, Charlottesville, VA 22903, USA}
   
   \date{Received 19 October 1999 / Accepted 26 January 2000}

   \maketitle

   \begin{abstract}

We present an inventory of mid-infrared spectral features detected in
high resolution (R$\sim$1500) ISO-SWS 2.4--45$\mu$m spectra of the 
galaxies \object{M\,82}, \object{NGC\,253}, \object{Circinus},
\object{NGC\,1068}, and a position in the \object{30\,Doradus} region of the 
Large Magellanic Cloud.
We discuss their identifications and highlight possible relations between
these features and the physical state of the interstellar medium in
galaxies. The spectral features vary considerably from source to source in 
presence and relative strength. Emission features are largely absent in 
the intense radiation field close to an AGN. Compared to normal 
infrared-selected starbursts, they also seem to be weaker in a
low metallicity, intensely star forming environment.
The large number of features beyond 13$\mu$m is remarkable. 
Some of the features 
have -- to our knowledge -- not been 
reported before in astronomical objects. 

In the 5--13$\mu$m region, emission from unidentified infrared bands (UIBs),
usually ascribed to
aromatic molecules, and apparent
silicate absorption dominate the spectrum. The density 
of features makes it
difficult to determine the continuum, particularly in ground-based data of 
limited wavelength coverage. In fact  
the apparent depth of the 9.7$\mu$m
silicate absorption may be overestimated in the presence of UIB emission, as
we demonstrate by comparing the spectrum of M\,82 to the (absorption free)
spectrum of the reflection nebula \object{NGC\,7023}. 
No strong silicate absorption is 
present in M\,82.
The (very small grain) dust continuum 
under the UIB emission in our starburst templates 
can be modeled by a simple power law, starting at wavelengths
between 8 and 9$\mu$m.

We find broad H$_2$O-ice absorption features at 3.0$\mu$m in M\,82 and 
NGC\,253. Their optical depths (relative to the visual extinction) 
indicate that the lines of sight towards these galaxies have similar 
properties as the line of sight
towards the Galactic Center.
The active galaxy NGC\,1068 exhibits a clearly different spectrum of 
absorption features, indicating different
physical conditions in the obscuring regions of this AGN compared to the
starburst templates.

The spectra are valuable templates for future
mid-infrared missions. 
We smooth our data to simulate low resolution spectra as obtained
with ISOCAM-CVF, ISOPHOT-S, and in the future with the low resolution mode of
SIRTF-IRS, and use our 
high
spectral resolution information to highlight possible identification problems
at low resolving power that are caused by coincidences of lines and features. 
The spectra are available in electronic form from the authors.

   \end{abstract}

%

\section{Introduction}
Mid-infrared spectra of galaxies are rich in emission lines, and
display prominent broader emission and absorption features due to the presence
of various solids and/or large molecules in their interstellar medium
(ISM). Significant variation from source to source suggests 
that these features may provide important
diagnostics of the ISM conditions in galaxies. 

The ground based and Kuiper Airborne Observatory spectra of the 
prototypical starburst M\,82 by Gillett et al. (1975) and Willner et al. (1977) 
fully established the existence of the
mid-infrared `unidentified infrared bands' (UIB) at 6.2, 7.7, 8.6, 
and 11.3$\mu$m
in galaxy spectra. These emission bands are characteristic of C-C and C-H bonds
in aromatic molecules. In this paper we will refer to them as 
`PAH features' 
according to one of the most popular interpretations of their carrier, 
polycyclic aromatic hydrocarbon molecules
\footnote{Other suggested carriers include small grains of
hydrogenated amorphous carbon (HACs), quenched carbonaceous composites (QCCs),
or coal.}. 
These detections and
related work using the IRAS LRS (Cohen \& Volk 1989) form the basic pre-ISO
knowledge of mid-infrared spectral features in galaxies.

Considerable work has also been
done from the ground but has been limited to the features
found in atmospheric windows, mainly silicate absorption and PAH feature
emission in the N band (e.g. Roche et al. 1991, and references therein) 
and the companion PAH feature
in the L band (e.g. Moorwood 1986).
The restriction
to atmospheric windows increases problems in establishing the `continuum' on
which the features are superposed. This is a nontrivial task, even with full
wavelength coverage, due to the crowding of mid-infrared 
emission and absorption features (especially in the 10$\mu$m region).

With the Short Wavelength Spectrometer SWS (de~Graauw et al. 1996) on
board
the Infrared Space Observatory ISO (Kessler et al. 1996) high spectral
resolution observations with good signal-to-noise (S/N) were
obtained for a number of bright galaxies. Their main
advantages lie in continuous wavelength coverage from 2.4 to 45$\mu$m
and in the possibility to clearly separate features from nearby
emission lines.

The interpretation of
galaxy-integrated spectra strongly benefits from comparisons to similar
observations of galactic
sources, sometimes spatially resolved, allowing
better isolation of the physical mechanisms at work. Recent ISO spectra
of many galactic template objects, such as reflection
nebulae (e.g. Boulanger et al. 1996, Cesarsky et al. 1996a, Verstraete et al.
1996, Moutou et al. 1998), planetary nebulae and circumstellar regions
(e.g. Beintema et al. 1996), and HII regions (Roelfsema et al. 1996,
Cesarsky et al. 1996b) have clearly demonstrated the importance of such
template spectra. They
prove that PAHs
are an ubiquitous part of the ISM. Additional information on emission features 
of crystalline silicates comes from similar template observations with ISO of 
e.g. planetary nebulae (Waters et al. 1998), evolved
stars (Waters et al. 1996), young stars (Waelkens et al. 1996), or LBVs in 
the LMC (Voors et al. 1999).
Absorption features (silicates, ices) have been found e.g. in the Galactic
center (Lutz et al. 1996, Chiar et al. 2000),  young stellar objects
(d'Hendecourt et al. 1996, Whittet et al. 1996, Dartois et al. 1999), and 
in dark clouds in the solar neighborhood (Whittet et al. 1998).

In this paper we present an inventory of mid-infrared spectral features 
detected in high resolution (R$\sim$1500) ISO-SWS 2.4--45$\mu$m spectra of the 
starburst galaxies M\,82 and NGC\,253, the Seyfert 2 galaxies 
Circinus and NGC\,1068, 
and a position in the 30\,Doradus star forming region of the Large Magellanic 
Cloud (Sect. \ref{s:inventory}). 
We briefly discuss possible feature identifications (Sect. \ref{s:ident}) 
and highlight possible relations 
between these features and the physical state of the interstellar medium in
galaxies (Sect. \ref{s:PAH_var}). 
We also address the issue of the continuum determination and the 
apparent depth of the silicate absorption at 9.7$\mu$m (Sect. 
\ref{s:continuum}). 
Finally (Sects. \ref{s:lowres} and \ref{s:conclusions}) we demonstrate
the use of these ISO spectra as templates for future 
infrared missions such as SIRTF, with particular emphasis on 
potential identification problems
at low resolving power that are caused by coincidences of lines and features.

All the spectra shown here exhibit a large number of atomic, ionic and
molecular emission lines. These have been or will be discussed elsewhere, 
along with more details on observations and data processing
(Circinus: Moorwood et al. 1996; M\,82: Lutz et al. 1998b, Schreiber 1998;
NGC\,1068: Lutz et al. 2000; 30\,Dor: Thornley et al., in prep.).  


\section{Observations and data reduction}

The objects discussed here have been observed as part of the ISO guaranteed 
time project on bright galactic nuclei. Here we concentrate on full grating 
scans obtained in SWS01 mode, speed 4. This mode provides a full 
2.4--45$\mu$m scan at resolving power of approximately 1000--2000.
For NGC\,253, Circinus, and NGC\,1068 the observations were centered on the
nuclei. In the case of M\,82 the observation was centered on the 
southwestern star formation 
lobe. For 30\,Dor the apertures were lying roughly parallel to an ionized
shell region, about 0.5\arcmin\/ away from the central stellar cluster.
Table \ref{tab:positions} summarizes the positions.
Note that different parts of an SWS full grating scan are observed with
different aperture sizes\footnote{The aperture sizes are
14\arcsec$\times$20\arcsec\/ for 2.4--12.0$\mu$m, 14\arcsec$\times$27\arcsec\/ 
for 12.0--27.5$\mu$m, 20\arcsec$\times$27\arcsec\/ for 27.5--29.0$\mu$m, and
20\arcsec$\times$33\arcsec\/ for 29.0--45$\mu$m, with some wavelength overlap
between the bands.}, varying between 14\arcsec$\times$20\arcsec\/ and 
20\arcsec$\times$33\arcsec.

We have processed the data using the SWS Interactive Analysis (IA) system
(Lahuis et al. 1998, Wieprecht et al. 1998) and the ISO Spectral Analysis
Package ISAP (Sturm et al. 1998). Dark current subtraction, scan direction 
matching, and flatfielding have been done interactively, and noisy 
detectors have been eliminated. In ISAP we clipped outliers and
averaged the data of all 12 detectors for each AOT band, retaining the 
instrumental resolution. For those wavelength ranges affected by fringes, the 
averaged spectra were defringed using the FFT or iterative sine fitting 
options of the defringe module within ISAP. 

\begin{table}
\caption[]{\label{tab:positions} Summary of observed positions (J2000), 
and position angles.}
\begin{flushleft}
\begin{tabular}{lllll}
\hline\noalign{\smallskip}
Object  &  RA  &  Decl. & PA & remark\\
\noalign{\smallskip}
\hline\noalign{\smallskip}
M\,82     &  09$^h$55$^m$50.7$^s$  &  69\degr40\arcmin44.4\arcsec & 245\degr & 1 \\
NGC\,253  &  00$^h$47$^m$33.2$^s$  & -25\degr17\arcmin17.2\arcsec &  28\degr & 2 \\
30\,Dor   &  05$^h$38$^m$46.0$^s$  & -69\degr05\arcmin07.9\arcsec & 230\degr & 3 \\
Circinus  &  14$^h$13$^m$09.7$^s$  & -65\degr20\arcmin21.5\arcsec &  19\degr & 2 \\
NGC\,1068 &  02$^h$42$^m$40.8$^s$  & -00\degr00\arcmin47.3\arcsec & -11\degr & 2 \\
\noalign{\smallskip}
\hline
\end{tabular}
\end{flushleft}
1 = SW lobe\\
2 = nucleus\\
3 = ionized shell
\end{table}

\begin{figure*}
 \setcounter{figure}{0}
 \resizebox{\hsize}{!}{\includegraphics{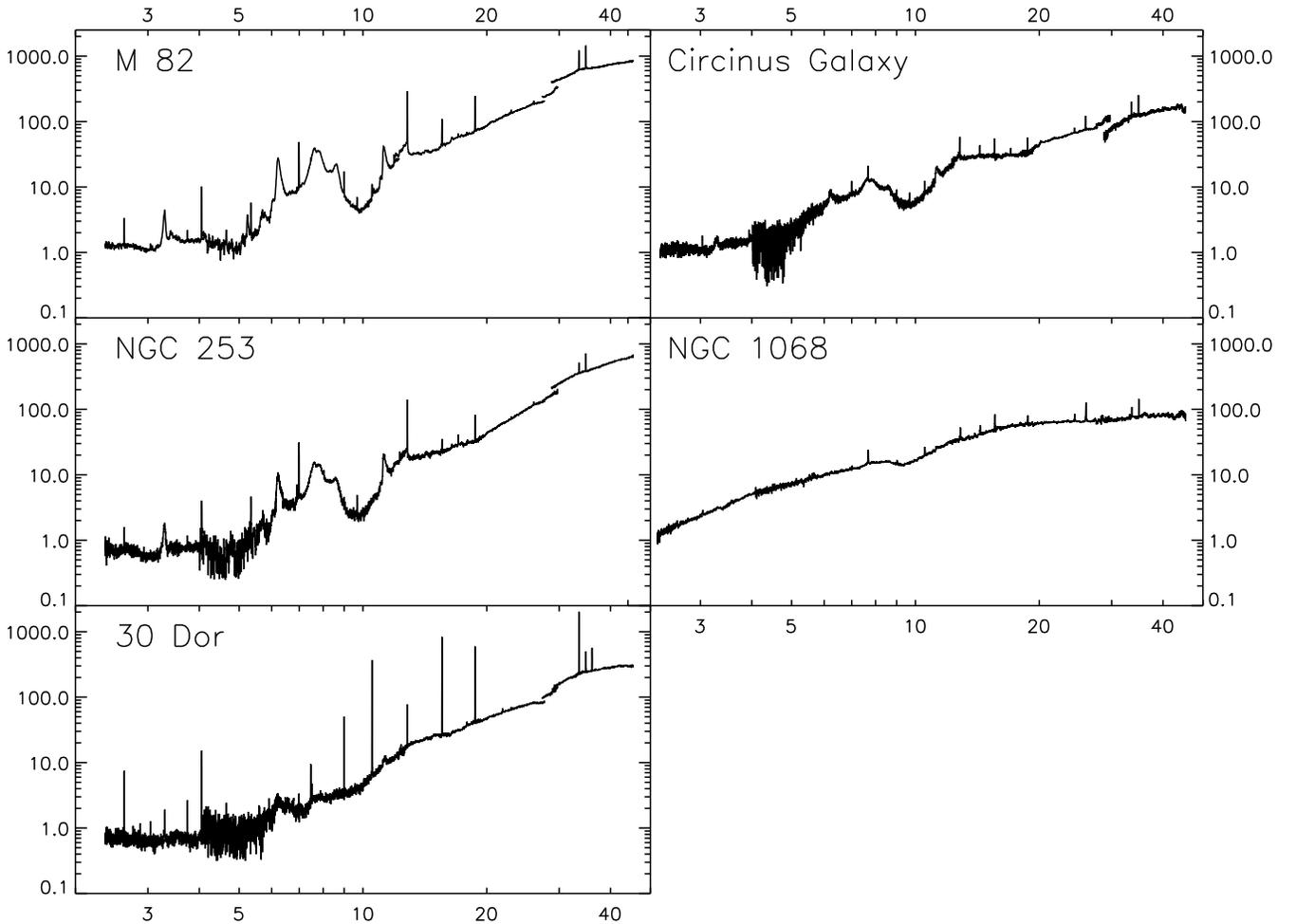}}
 \caption{The full SWS01 spectra of the five extragalactic templates. x-axis:
  wavelength in $\mu$m; y-axis: flux density in Jansky.}
 \label{fig:m82}
\end{figure*} 

To reduce noise for display purposes (Fig. 1), we have smoothed the data 
with a gaussian filter to a uniform resolution of 1000. This
broadens slightly the line widths of the narrow atomic and ionic emission
lines, but does not affect the broad emission and
absorption features discussed in this paper.
We did not remove the flux jumps at detector band limits. Part of
these jumps may be real because the sources are extended and because
the aperture sizes
change at some band edges (at approximately 12.0, 27.5 and 29.0$\mu$m).
Another part simply reflects the flux calibration
uncertainty, which is of the order of 20 per cent.

We carefully checked the reality of the features in the final spectra 
against the possibility of residual instrumental features from the 
Relative Spectral Response Function (RSRF), which might be caused 
e.g. by an improper dark current subtraction.
For example, the detector RSRF exhibits absorption features 
at 11.05 and 34$\mu$m
which might appear in emission in the calibrated spectrum. This is, however,
at most an effect of the order of a few per cent of the continuum level. 
The features
we see at these wavelengths in our spectra (see below) are stronger, so
that they must be real.
Furthermore, we checked whether a feature
appears in both scan directions and in the majority of all
detectors. Additional confirmation
was possible for those features that lie in the overlap
region of two different AOT bands and appear in both bands.
At this stage of instrument calibration, we do not believe that any
of the broad structures in band 3E (appr. 27.5--29.5 $\mu$m) are real
features.
Also, features at the end of band 4 (43--45 $\mu$m, e.g. in Circinus)
cannot be trusted.


\section{An inventory of features}
\label{s:inventory}

\begin{table}
\caption[]{\label{tab:inventory} Summary of observed broad emission features
(approximate peak positions in $\mu$m). Uncertain detections are given in
parentheses, nondetections are marked by a dash.}                    
\begin{flushleft}
\begin{tabular}{llllll}
\hline\noalign{\smallskip}
M\,82  &  NGC  &  30\,Dor  & Circinus  &  NGC   & nearby\\
       &  253  &           &           &  1068  & lines\\
\noalign{\smallskip}
\hline\noalign{\smallskip}
3.25 & 3.25  &  --    & 3.25  & --    & \\
3.3  & 3.3   & 3.3    & 3.3   & 3.3   & Pf$_{\delta}$\\
3.4  & 3.4   & 3.4    & --    & --    & \\
3.5  & (3.5) & 3.5    & --    & --    & \\
--   & 3.75  &  --    & 3.75  & --    & \\
5.25 &  --   &  --    & --    & --    & \\
5.65 & 5.65  &  --    & --    & --    & \\
6.0  & 6.0   & 6.0    & (6.0) & --    & \\
6.2  & 6.2   & 6.2    & 6.2   & (6.2) & \\
6.3  & 6.3   & 6.3    & 6.3   & --    & \\
7.0  & 7.0   & (7.0)  & --    & --    & H$_2$+\\
     &       &        &       &       & [Ar II]\\
7.6  & 7.6   & 7.6    & 7.6   &(7.6)  & Pf$_{\alpha}$+\\
     &       &        &       &       & [Ne VI]\\
7.8  & 7.8   & 7.8    & 7.8   & 7.8   & \\
8.3  & --    & --     & --    & --    & \\
8.6  & 8.6   & 8.6    & 8.6   & 8.6   & \\
(10.6)& --   & --     & --    & --    & [S IV]\\
11.05 & 11.05& 11.05  & 11.05 & 11.05 & \\
11.25 & 11.2 & 11.25  & 11.25 & 11.25 & \\
12.0  & 12.0 & 12.0   & --    & --    & \\
12.7  & 12.7 & --     & 12.7  & (12.7)& [Ne II]\\
13.55 & 13.55& (13.5) & 13.6  & 13.4  & \\
14.25 & 14.25& 14.25  & 14.25 & --    & [Ne V]+\\
      &      &        &       &       & [Cl II]\\
(14.8)&(14.8)&(14.8)  &(14.8) & --    & \\
15.7  & 15.7 & 15.7   & 15.85 & 15.9  & [Ne III]\\
16.5  & 16.5 & 16.5   &  --   & --    & \\
(17.4)&(17.4)& --     & --    & --    & \\
(18.0)& --   &(18.0)  &  --   & --    & \\
20.5  &(20.3)&(20.4)  & 20.2  & --    & \\
--    & --   & --     & 21.7  & --    & \\
34    & 34   &  34    & 34    &  --   & [Si II]+\\
     &       &        &       &       & [S III]\\
%
\noalign{\smallskip}
\hline
\end{tabular}
\end{flushleft}
\end{table}

\begin{table}
\caption[]{\label{tab:invent_abs} Summary of observed absorption features.
Uncertain detections are given in parentheses.}                    
\begin{flushleft}
\begin{tabular}{llllll}
\hline\noalign{\smallskip}
M\,82  &  NGC  &  30\,Dor  & Circinus  &  NGC   & species\\
       &  253  &           &           &  1068  &        \\
\noalign{\smallskip}
\hline\noalign{\smallskip}
3.0    & 3.0   & --    & --     & --  & H$_2$O ice  \\
--$^1$ & --$^1$& --$^1$& --$^1$ & 3.4 & hydrocarbon \\
(9.7)  & (9.7) &  --   & (9.7)  & 9.4 & silicate \\
(18.0 )& (18.0)&  --   & (18.0) &  -- & silicate \\
\noalign{\smallskip}
\hline
\end{tabular}
\end{flushleft}
$^1$ could be filled in by PAH emission
\end{table}

\begin{table}
\caption[]{\label{tab:PAH_flux1} Approximate absolute and relative fluxes of the 
prominent PAHs at 3.3 and 6.2$\mu$m (see text for details).}
\begin{flushleft}
\begin{tabular}{lllll}
\hline\noalign{\smallskip}
Object   & \multicolumn{2}{c}{3.3$\mu$m}  & \multicolumn{2}{c}{6.2$\mu$m} \\ 
         & flux$^1$ & relative flux$^2$ & flux$^1$ & relative flux$^2$ \\
\noalign{\smallskip}
\hline\noalign{\smallskip}
M\,82     & 3.3 & 0.2 & 37  & 2.7 \\
NGC\,253  & 1.0 & 0.1 & 13  & 1.7 \\
30\,Dor   & 0.3 & 0.04& 1.5 & 0.2 \\
Circinus  & 0.7 & 0.06& 6.3 & 0.6 \\
NGC\,1068 & 0.3 & 0.02& 1.0 & 0.05\\
\noalign{\smallskip}
\hline
\end{tabular}
\end{flushleft}
$^1$ [10$^{-18}$ W/cm$^2$] \\
$^2$ PAH flux/continuum(11.6--11.9$\mu$m)
\end{table}

\begin{table}
\caption[]{\label{tab:PAH_flux2} Approximate absolute and relative fluxes of the 
prominent PAHs at 7.7 and 11.3$\mu$m (see text for details).}
\begin{flushleft}
\begin{tabular}{lllll}
\hline\noalign{\smallskip}
Object   & \multicolumn{2}{c}{7.6--7.8$\mu$m}  & \multicolumn{2}{c}{11.3$\mu$m} \\ 
         & flux$^1$ & relative flux$^2$ & flux$^1$ & relative flux$^2$ \\
\noalign{\smallskip}
\hline\noalign{\smallskip}
M\,82     & 99  & 7.2 & 16  & 1.2 \\
NGC\,253  & 39  & 5.0 & 7.2 & 0.9 \\
30\,Dor   & 0.3 & 0.04& 1.7 & 0.2 \\
Circinus  & 21  & 1.9 & 3.9 & 0.4 \\
NGC\,1068 & 2.1 & 0.1 & 1.1 & 0.06\\
\noalign{\smallskip}
\hline
\end{tabular}
\end{flushleft}
$^1$ [10$^{-18}$ W/cm$^2$] \\
$^2$ PAH flux/continuum(11.6--11.9$\mu$m)
\end{table}

In Tables \ref{tab:inventory} and \ref{tab:invent_abs} 
we give an inventory of features that we believe to be reliable detections. 
We consider a detection reliable if the feature has an amplitude of at least
3$\sigma$ of the noise level and if it fulfills the criteria described above.
We also list a few uncertain detections in parantheses, e.g. features in
compliance with the above criteria but with an amplitude of less than
3$\sigma$ (but note that the definition of the local noise level is in many 
cases somewhat uncertain).
Most of these features have been described before in reports of ISO-SWS
observations of galactic template sources (e.g. Moutou et al. 1996, 
Verstraete et al. 1996, 
Roelfsema et al. 1996, Beintema et al. 1996). Here we highlight  
a few of their characteristics, source by source. In Sect. \ref{s:ident}
we briefly discuss possible identifications. Please note that many of 
the features are severely blended and thus the peak wavelengths given here are 
approximate. 

We also list the fluxes of four of the most prominent PAHs in
Tables \ref{tab:PAH_flux1} and \ref{tab:PAH_flux2}. To measure these fluxes
we defined continua by a linear interpolation between the following points: 
2.50 and 3.65$\mu$m for the 3.3$\mu$m feature, 5.9 and 10.9$\mu$m for the
features at 6.2 and 7.7$\mu$m, and 10.9 and 11.8$\mu$m for the 11.3$\mu$m
feature. We then obtained the fluxes by integrating between the following band
limits: 3.10--3.35$\mu$m, 6.0--6.5$\mu$m, 7.3--8.2$\mu$m, and
11.1--11.7$\mu$m. To give an indication of the relative contribution of these
features to the infrared luminosity we also list their ratio to the continuum
flux in the range 11.6--11.9$\mu$m. Due to the uncertainties involved in this
measuring process (continuum shape, feature profile, etc.) 
all absolute and relative fluxes are only approximate.
By definition, these fluxes ignore a possible, PAH-related `plateau' or
`continuum' in the 6--9 and 10--13$\mu$m range (e.g. Boulanger et al. 1996).\\

\begin{figure*}
\setcounter{figure}{1}
 \resizebox{\hsize}{!}{\includegraphics{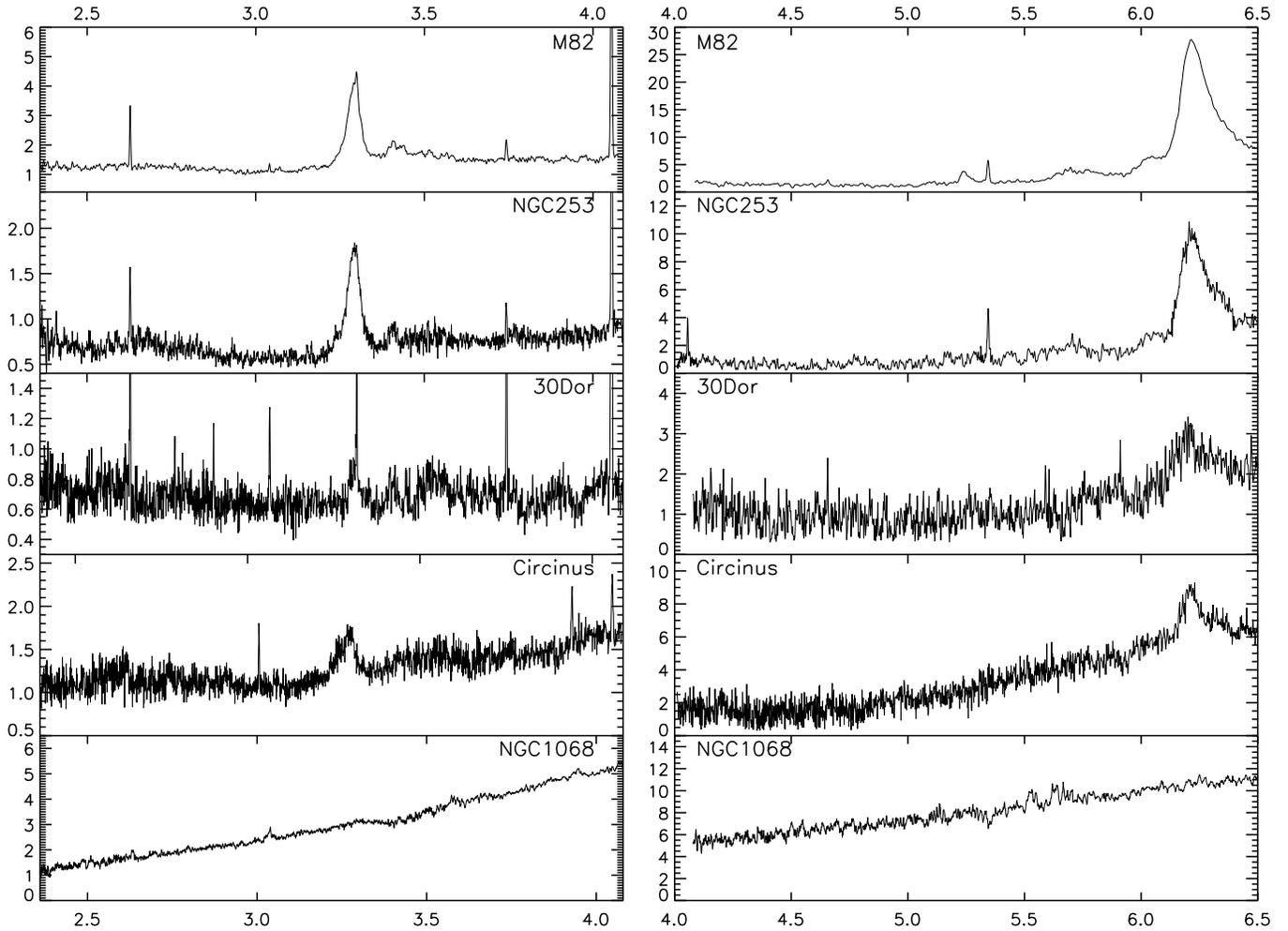}}
 \vspace*{0.05cm} \caption{Details of the spectra. Wavelength in $\mu$m, flux 
  density in
 Jansky.}
 \label{fig:details}
\end{figure*} 
\begin{figure*}
\setcounter{figure}{1}
 \resizebox{\hsize}{!}{\includegraphics{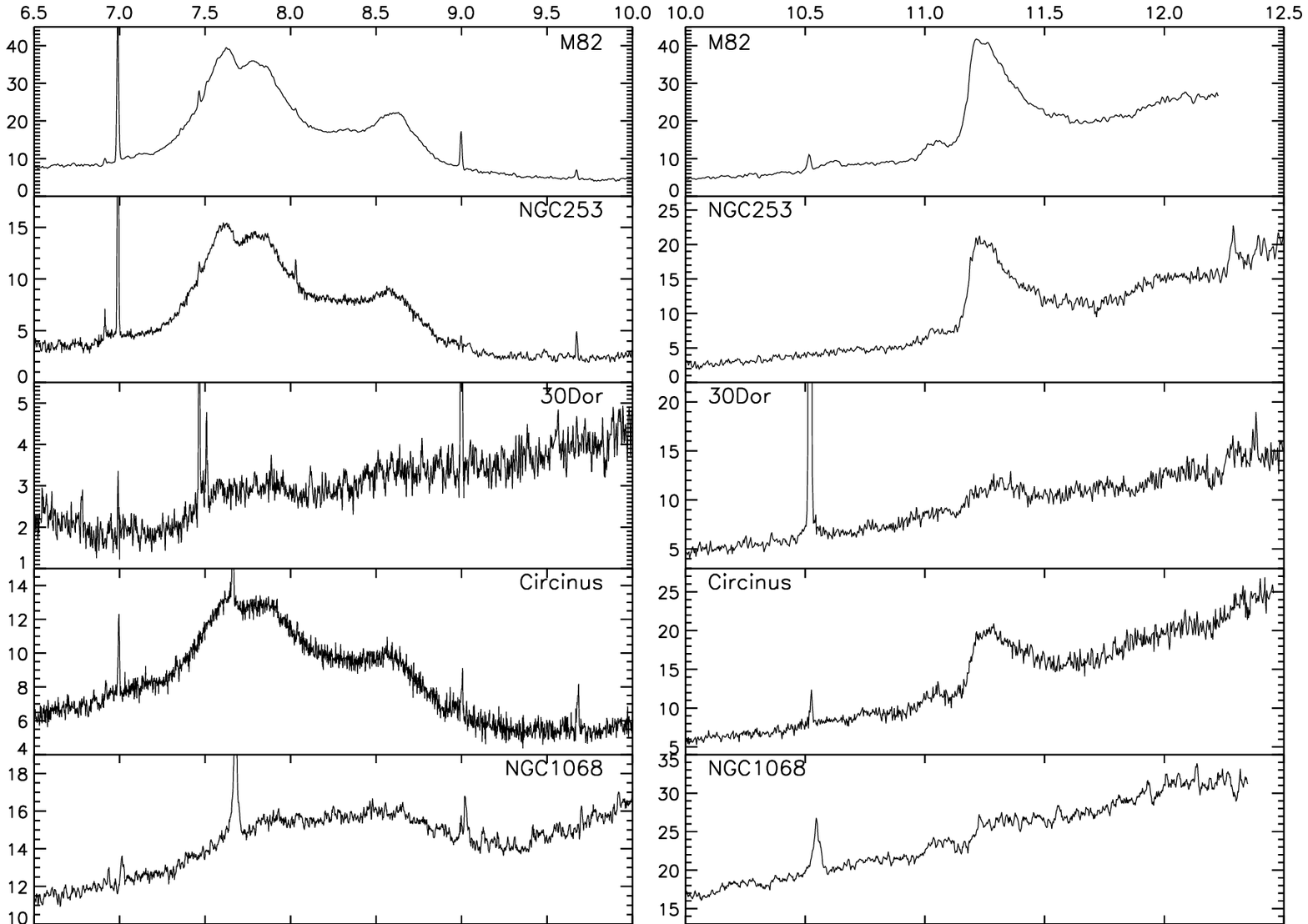}}
 \vspace*{0.05cm} \caption{{\it continued} }
\end{figure*} 

{\noindent \bf M\,82:} 
M\,82 is a small galaxy undergoing a very powerful
starburst, and is considered to be
a prototype of starburst activity.
Due to its proximity (3.63 Mpc, Freedman et al. 1994) it is the
brightest galaxy in the infrared, with the infrared luminosity arising mainly
from warm dust in the central region.
Emission features
at 8.7 and 11.3$\mu$m were first detected by Gillett et al (1975), and the
3.3, 6.2 and 7.6$\mu$m features by Willner et al. (1977).

The main PAH features are clearly seen in the
mid-infrared spectrum of M\,82, shown in its entirety in Fig. 1
and in detail in the top panels of Fig 2. In addition,
the spectrum shows a large number of weaker features
which have previously been detected with ISO only in
galactic template sources.

The 3.3$\mu$m feature has satellites at 3.4 and 3.5$\mu$m.
It also shows a blue asymmetry, which might be due to a feature at 3.25$\mu$m
(see Beintema et al. 1996).
Two weak features at 5.25 and 5.65$\mu$m, that have been observed e.g. in
the planetary nebula NGC\,7027 (Beintema et al. 1996) and the
photodissociation front of M\,17 (Verstraete et al. 1996), are also present
in M\,82.
The 6.2$\mu$m band has a red shoulder  and
shows an additional feature at 6.0$\mu$m.
The 7.7$\mu$m feature consists of two bands at 7.6 and 7.8$\mu$m.
A significant contribution of 7.7$\mu$m solid methane absorption 
(e.g. Whittet et al. 1996, Lutz et al. 1996) to this strong dip between 
7.6 and 7.8$\mu$m is unlikely, since in M\,82 related icy absorption features 
are shallower and extinction is lower (Sect. \ref{s:continuum}) than in the 
sources with clear methane absorptions.
A weak feature is present at 10.6$\mu$m, which is confirmed
in the average starburst spectrum of Lutz et al. 1998a
(their Fig. 1, independent ISOPHOT-S data), and probably related
to a feature seen by Beintema et al. (1996) in the spectrum of NGC\,7027. 
The 11.3$\mu$m feature peaks at 11.2$\mu$m and shows the
well-known asymmetric shape towards longer wavelengths (Witteborn et al.
1989).
There is an additional component around 11.05$\mu$m, much too strong to be
related to artifacts from the RSRF correction known to exist at this wavelength.
A weak feature may be present near 12.0$\mu$m. 
The 12.7$\mu$m feature is very prominent in M\,82.
Moutou et al. (1998) found two emission features at 15.8 and 16.4$\mu$m in
the spectrum of the galactic reflection nebula NGC\,7023. On the basis
of their laboratory work,  they
attributed these features to PAH molecules.
These two features are clearly seen in M\,82, and in some of our other 
template spectra.
More emission features can be found at still longer wavelenghts
(20.5 and 33-34$\mu$m).

In addition we find two features that -- to our knowledge --
have not been reported before: at 7.0 and 8.3$\mu$m.
A feature around 7.0$\mu$m is also present in NGC\,253 (most clearly), and 
perhaps in 30\,Dor.
We found the same feature in ISO-SWS 
spectra of the cool, dusty envelopes of the planetary nebula He\,2-113
(see e.g. Waters et al. 1998, their Fig. 2).
The average ISOPHOT-S spectra of starburst and normal galaxies (Lutz et al.
1998a, Helou et al. 2000) also show hints of a weak feature, blended with
H$_2$ S(5) 6.91$\mu$m and [Ar II] 6.99$\mu$m (see also Sect. \ref{s:lowres}). 
The 8.3 feature is also visible in
the galactic template spectrum of NGC\,7023 (Fig. \ref{fig:m82vsngc7023}),
and perhaps in some of the compact HII regions shown in Roelfsema et al. (1996).
We also want to mention here the 14.3$\mu$m band. An astronomical observation 
of this band has been reported only recently for the first time 
(Tielens et al. 1999). It is present in all 
our template spectra, with the
exception of NGC\,1068, in NGC\,7023, and perhaps also in some of
the circumstellar PAH spectra shown in Beintema et al. (1996).
On the other hand, our spectra do not show some of the 
emission features that have been detected before 
in astronomical observations,
e.g. at 4.65$\mu$m (Verstraete et al. 1996) and 
13.3$\mu$m (Moutou et al. 1998).

\begin{figure*}
\setcounter{figure}{1}
 \resizebox{\hsize}{!}{\includegraphics{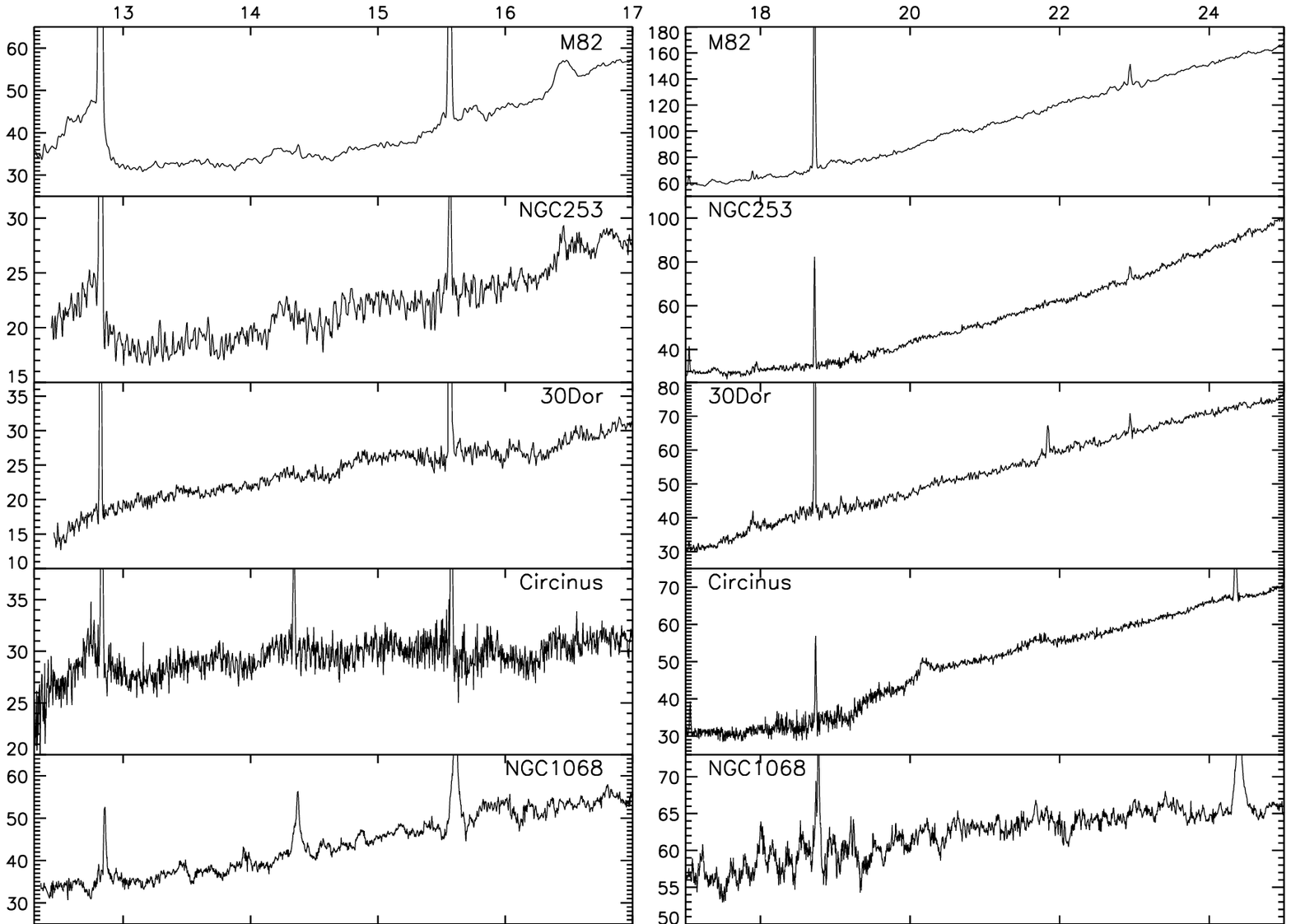}}
 \vspace*{0.05cm} \caption{{\it continued} }
\end{figure*} 
\begin{figure*}
\setcounter{figure}{1}
 \resizebox{\hsize}{!}{\includegraphics{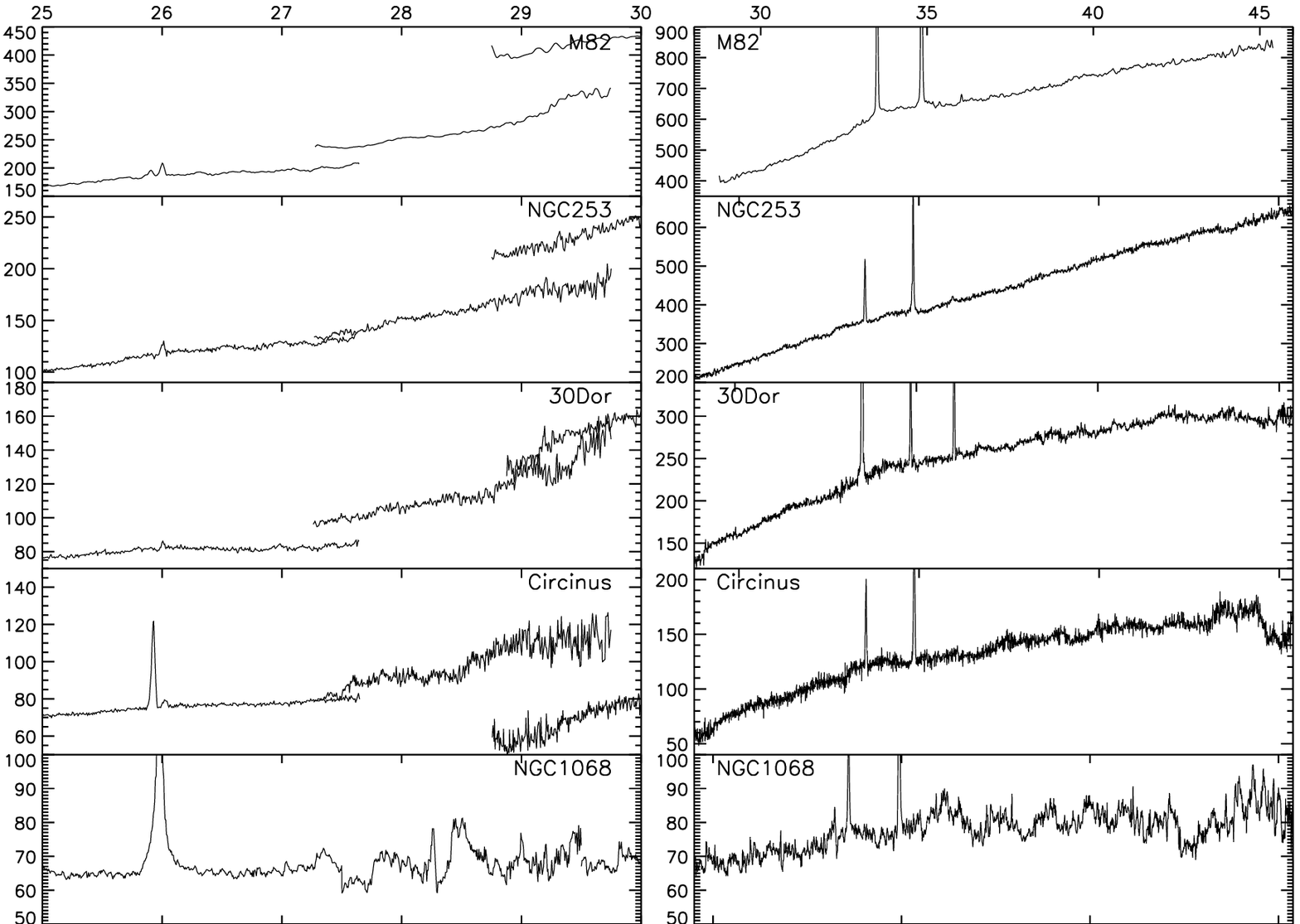}}
 \vspace*{0.05cm} \caption{{\it continued}. Broad features between 27.5 and 
                29.5$\mu$m, and longward of 40$\mu$m cannot be trusted.}
\end{figure*} 

There are very few absorption features in our spectrum of M\,82. The
trough around 10$\mu$m looks like a strong 9.7$\mu$m silicate feature.
In
Sect. \ref{s:continuum} we argue, however, that the trough is mainly due 
to the strong PAH emission at 8.7 and 11.3$\mu$m. 
Also, there is no clear signature 
of a corresponding silicate absorption feature at 18$\mu$m.
A broad absorption feature around 3.0$\mu$m is probably due to H$_2$O-ice.
Its optical depth ($\tau\sim$0.2) is relatively small 
compared e.g. to the Galactic center ($\tau$=0.5, Lutz et al. 1996, 
Chiar et al. 2000). However, 
since the overall extinction estimates differ in the same sense,
this is consistent with the M\,82 line of sight having properties similar to that
towards the Galactic center: a mixture of diffuse ISM and molecular cloud
extinction, with some 
variance in the relative weight for different lines of sight
(Chiar et al. 2000). These properties seem to be quite typical for 
starburst galaxies, as we find similar conditions in NGC\,253 (see below) 
and in NGC\,4945 (Spoon et al., in prep.).

The M\,82 spectrum has the highest S/N ratio and
shows the largest number of features in our sample.
Combined with the ISO-LWS long wavelength spectrum (Colbert
et al. 1999) it will be an important template for future missions.\\

{\noindent \bf NGC\,253:}
NGC\,253 is a nearby, almost edge-on barred spiral galaxy with a high level 
of circumnuclear 
starburst activity. At optical wavelengths the galaxy is heavily obscured by
dust lanes in the central regions.
The ionic emission lines in NGC\,253 are of lower excitation (e.g. lower
[Ne III]/[Ne II] ratio) than in M\,82, suggesting a softer average
radiation field (Thornley et al., in prep.). 
Despite these differences between the two galaxies, their 
spectra of broad emission features are 
remarkably similar. This is demonstrated in Fig. \ref{fig:m82vsngc253}.
Only beyond 9$\mu$m does a difference in the underlying continuum 
become obvious. NGC\,253
has a stronger continuum at these longer wavelengths. 
In starburst galaxies this underlying continuum is usually attributed to 
very small grains (VSG) of dust (e.g. D\'esert et al. 1990). 
In Sect. \ref{s:continuum} we will use the difference in the continua of
M\,82 and NGC\,253 to characterise the shape of the VSG continuum.

Table \ref{tab:inventory} shows that nearly all features of M\,82 are also seen
in NGC\,253. The very few exceptions may simply be due to the lower
S/N ratio of the NGC\,253 spectrum.
The red shoulder of the 6.2$\mu$m band seen in M\,82
is more prominent in NGC\,253 and is probably due to an additional feature at
6.35$\mu$m.
The 3.0$\mu$m ice absorption is also present, having an optical depth of
$\tau\sim$0.25.\\

{\noindent \bf 30\,Dor:}
The 30\,Dor region in the LMC is the largest, most
massive, and most luminous H II region in the Local Group.
As a local template for massive star
formation and its interaction with the interstellar medium, it is
instructive to compare its spectrum to the galaxy spectra  in
our sample.
A more detailed study of the mid-infrared fine structure emission lines in
30\,Dor will be presented in an upcoming paper (Thornley et al., in prep.).

An inspection of Fig. \ref{fig:details} and Table \ref{tab:inventory} shows
that the 30\,Dor spectrum exhibits most of the PAH features found in the
galaxy spectra. Lower S/N may contribute to
the non-detection of some of the weak features.
Compared to the two starburst galaxies, the features in 30\,Dor are
much weaker relative to the continuum (see also Tables \ref{tab:PAH_flux1}
and \ref{tab:PAH_flux2}) and show different ratios.
Note, for instance, the high 6.2/7.7$\mu$m feature ratio, 
the unusually high 3.4/3.3 ratio ($\approx$ 0.5, see also Sect. 
\ref{s:PAH_var}), the shape of the 7.6/7.8$\mu$m features, or the complete
absence of the 12.7$\mu$m feature.
These differences may be partly due to the fact, that the SWS apertures 
covered only part of the entire 30\,Dor complex. Verstraete et al. (1996)
use the case of M17 to demonstrate the strong variations in PAH spectra going
from the center of an H\,II region to the surrounding photodissociation region. 
On the other hand, the similarity of the 30\, Dor 
spectrum to the integrated ISOPHOT-S spectrum of the dwarf galaxy NGC\,5253 
(Rigopoulou et al. 1999) strongly suggests that the weakness of the PAHs is not
an aperture effect but reflecting an intrinsic property of very active
star formation in a low metallicity environment.

Silicate absorption at 9.7 and 18$\mu$m is, if at all present, very weak.
No other absorption features can be detected. \\

{\noindent \bf NGC\,1068:}
The nearby, prototypical Seyfert 2 galaxy NGC\,1068 is a key object in
the investigation and modeling of active galactic nuclei (AGNs).
The ISO-SWS observations were centered on the active nucleus, with the
aperture
covering very little of the
circumnuclear star forming `ring', which has a radius of $\approx$15\arcsec.
In stark contrast to the starburst templates we have shown, this active
nucleus spectrum shows very little PAH emission.
The weaker features (e.g. 3.3, 6.2 or 12.7$\mu$m)
are barely visible, if at all. 

The continuum around 10$\mu$m is strong because of a central warm component
heated by the AGN. Silicate absorption is clearly present, although centered
at 9.4$\mu$m rather than 
at 9.7$\mu$m, but again surrounding weak PAH emission complicates its
interpretation. 
Hydrocarbon absorption is seen at 3.4$\mu$m (see also
Bridger et al. 1994), in contrast to our starburst templates, where
it is not observed.
Note, however, that in M\,82 and NGC\,253 a similar absorption feature
could plausibly
exist if the analogy to the Galactic center holds, but be filled in by 
the 3.3/3.4 PAH emission. On the other hand
M\,82 and NGC\,253 show an H$_2$O ice absorption at 3.0$\mu$m which is
definitely absent in NGC\,1068.  
These differences in absorption features are entirely plausible because of the
different physical conditions in the obscuring regions. For the starbursts,
they likely include diffuse ISM as well as molecular clouds that can host icy
grains. Conversely, infrared polarimetry suggests that most of the 
near-infrared obscuration in NGC\,1068 occurs within a few parsecs from the 
nucleus, possibly in the torus (e.g. Packham et al. 1997). Such an energetic 
environment will be much less favourable for the existence of icy grains.\\

{\noindent \bf Circinus:}
The Circinus Galaxy is a nearby spiral galaxy which shows Seyfert 2 activity
(e.g. Moorwood \& Glass 1984, Oliva et al. 1994, Moorwood et al. 1996,
Oliva et al. 1998). Due to its proximity (5 times closer than NGC\,1068)
it has become another template object for the study of AGNs.
The AGN is surrounded by circumnuclear
star-forming regions, as is often the case in Seyfert nuclei residing in
spirals.
The ISO-SWS observations were centered on the active nucleus, but contrary to
the observations of NGC\,1068 the apertures covered a significant amount of
this circumnuclear star formation.
Hence, most of the dust emission features seen in
the starburst templates are also found in
Circinus, but with weaker line-to-continuum ratio (see the discussion in
Sect. \ref{s:PAH_var}). A peculiarity of the Circinus spectrum are the
very pronounced features in the 20--22$\mu$m region.

The observation was performed very early in the mission, when the
observing strategy was not yet fully optimized. For instance, exposures of
the internal flux calibration lamps, preceeding observations of the scientific
target, caused memory effects in the immediately following scans.
The low flux level of band 4 ($\lambda \ge$ 29$\mu$m),
relative to the preceding bands, is due to such a memory effect and an incorrect
dark current subtraction. The same might be true for the apparent features near
44$\mu$m (which are not visible in the overlapping LWS spectrum). We did not 
attempt to improve the dark current subtraction further since this would 
involve subjective assumptions about the true dark current.

\section{Mid-infrared features and the physical state of the ISM}


\subsection{Identification}
\label{s:ident}

\subsubsection{The 2-13$\mu$m region:}

The emission features in this wavelength range
have been extensively studied with ISO in galactic objects during the last few
years. Comprehensive discussions of their identifications and characteristics
can be found e.g. in Beintema et al. (1996), Moutou et al. (1998), Roelfsema
et al. (1996), and  Verstraete et al. (1996).
They are most often attributed to PAH molecules. This is supported by
recent laboratory studies (e.g. Roelfsema et al. 1996, Moutou et al. 1996).
The only features in this range, 
which have not
been addressed in the
literature so far, are the ones at 7.0 and 8.3$\mu$m. 
In our sample of galaxies 
they are
unambiguously detected only in M\,82 and NGC\,253 (the 8.3$\mu$m feature only
in M\,82), but as mentioned in Sect.
\ref{s:inventory} they seem to be present in other published spectra as well.
It seems likely that they can be attributed to a PAH modes, too.

\subsubsection{Features between 13 and 20$\mu$m:}

Features in this range have attracted less attention in the past, because
they are intrinsically weak (but see e.g. Beintema et al. 1996).
These bands, however, are more sensitive to the molecular structure of PAHs,
since they
involve the motion of the molecule as a whole, therefore depending on the 
exact species (L\'eger et al. 1989).
Hence, their observation could help to better
constrain the composition of the interstellar mixture.
The 13.6, 15.8, and 16.5$\mu$m features are
also visible in the spectra of NGC\,7023 and have been attributed to PAH bands
in the past by Moutou et al. (1998), based on their laboratory work.
The band at 14.3$\mu$m has been tentatively attributed to a phenyl bending 
mode by Tielens et al. (1999). 
Moutou et al. (1996) list a feature at this wavelength in their composite
laboratory spectrum of a mixture of PAHs. Although weak, it might cause
confusion with [Ne V] in low resolution spectra (see Sect. \ref{s:lowres}).
Finally, the weak feature at 14.8$\mu$m (if real) could be due to the smallest 
PAH, benzene (Tielens et al. 1999).

\subsubsection{Features in the 20 to 45 $\mu$m region:}

The number of modes in the laboratory PAH spectra of Moutou et al. (1996) 
decreases
with increasing wavelength. Few species show emission beyond 20$\mu$m, 
e.g. near 21, 28 and 40$\mu$m.
In this wavelength range other sources must be taken into
account. A number of recent papers have reported the detection of crystalline
silicates (olivines, fosterite, pyroxene, etc.) in objects like 
Luminous Blue Variables (Voors et al. 1999),
dusty circumstellar disks (Waelkens et al. 1996, Waters et al. 1996), or
Planetary Nebulae (Waters et al. 1998).
In particular the feature at 34$\mu$m
could be attributed to
these kind of sources. However, one would expect to see emission
features at e.g. 23, 28, 40 and 43$\mu$m, as well; none of these features are
clearly detected in our spectra.
On the other hand, even in some of the galactic templates, such as the
planetary nebula NGC\,6543 (e.g. Waters et al. 1996), not all of
these features are present.

A feature near 20.5$\mu$m shows a striking variation in shape and central
wavelength between the galaxies. This is particularly surprising since 
the galaxy spectra include a mixture of many different regions, and may suggest
a carrier occuring only transiently in very special conditions.
In Circinus this feature is most prominent, and peaks at the bluest wavelength 
(20.2$\mu$m). Circinus also shows a second peak at 21.7$\mu$m which 
is absent in the other spectra. 
A bump around 20.5$\mu$m is also seen in ISO-SWS spectra of M supergiants 
(Voors et al. 1999, Molster et al. 1999). 
It could be due to PAHs or alternatively metal
oxides like FeO (Waters et al. 1996, Henning et al. 1995).
IRAS-LRS and ISO-SWS spectroscopy have also detected a broad feature, centered
at approximately 20.1$\mu$m, in carbon rich stars (Volk et al. 1999,
Garc\'{\i}a-Lario et al. 1999, Szczerba et al. 1999). Possible candidates 
that have
been proposed include large PAH clusters or hydrogenated amorphous carbon
grains, hydrogenated fullerenes, and nano-diamonds (see references in
Volk et al. 1999). However, compared to these detections, the
features
in our spectra are much narrower.

A mixture of fullerene molecules of different degree of hydrogenation (Webster
1995) might also explain the second peak in Circinus around 21.7$\mu$m, 
since the emission
peak shifts from 23 for fully hydrogenated fullerene (C$_{60}$H$_{60}$)
to 19$\mu$m for non-hydrogentated fullerene (C$_{60}$).
None of the other galaxies exhibit this feature, but
the SWS01 spectrum of NML Cyg (Voors et al. 1999) and of the galaxy
NGC4945 (S. Lord, private communication) also
show a weak emission feature around 21.6--22$\mu$m.

The broad plateau at 33--34$\mu$m, i.e. under the strong lines of [Si II]
and [S III], could be affected by detector memory effects. To remove such a
possible instrumental effect we treated the two
different scan directions of the SWS01 mode separately. 
The trailing wings of each line profile, i.e. the blue
wing for the scan with increasing wavelength, and the red wing for the scan
decreasing in wavelength, are much more distorted by memory effects than
the leading wings. Therefore, we cut out these trailing wings, before we
averaged the spectra of the two scan directions.
We are
hence confident that most of the remaining plateau is real. Such a feature
has been observed in many
galactic targets and is generally attributed to crystalline silicates
(olivine, e.g. Waters et al. 1998).



\subsection{Variation of PAH features}
\label{s:PAH_var}

Published mid-IR spectra of galactic template sources show a significant
variation of intrinsic
PAH ratios from source to source. For instance Roelfsema et al.
(1996) see a drastic change in the relative intensities of the 7.7 and 8.6
bands with increasing intensity or hardness of the radiation field.
Similar changes are seen e.g. in different regions of M17
(Verstraete et al. 1996) or - for the 8.6/11.3 ratio - in the
reflection nebula NGC\,1333 (Joblin et al. 1996).
PAHs exposed to intense and hard radiation fields can be ionized, lose
hydrogen atoms, or be photodissociated; any of these effects may contribute to
the observed variations in PAH ratios.
According to Joblin et al. (1996) the ionization is best traced by the 3.4/3.3
and 8.6/11.3 ratios. A good hydrogenation indicator is the (12+12.7)/11.3
ratio. For instance, these authors find a high 3.4/3.3 ratio 
of 0.1 in radiation fields that are 10$^5$ times the standard.

Another important factor that can alter observed
PAH ratios is extinction. Extinction
will suppress the 6.2, 8.6 and 11.3$\mu$m features with respect to 
the one at 7.7$\mu$m. 
The 12.7/11.3 ratio is similarly affected, since the 11.3 feature is
still in the wing of the 9.7$\mu$m silicate absorption. Details will depend
on the applicable extinction law (see e.g. the Galactic center, Lutz et al. 
1996). While extinction
clearly affects PAH spectra in highly obscured sources like Ultraluminous 
Infrared Galaxies (ULIRGs, Lutz et al. 1998a) or the edge-on galaxy 
NGC\,4945 (Spoon et al., in prep.), its
effect will be less pronounced in the lower extinction sources of our sample.

Finally, in active galaxies, such as NGC\,1068 or the Circinus Galaxy, PAH
features can be diluted by an AGN-powered hot dust continuum.
Genzel et al. (1998) and Lutz et al. (1998a) have used this as a diagnostic 
of the power sources of ULIRGs.

Our sample of galaxy spectra exhibits a similar trend in relative PAH
strengths as the galactic templates.
M\,82 and NGC\,253 have high
3.4/3.3 ratios, consistent with them being active starburst galaxies.
30\,Dor seems to have an even higher 3.4/3.3 ratio, but the S/N ratio is not
sufficient for a detailed analysis. However, a strong and hard, highly ionizing
radiation field in 30\,Dor is consistent with the results of Thornley et
al. (1998), which are based on the ratios of fine structure emission lines,
like [Ne III]/[Ne II]. In that context it is interesting to note again
the complete absence of the 12.7$\mu$m feature in 30\,Dor.
In the two starburst galaxies
M\,82 and NGC\,253 we see well-separated 7.6/7.8 and 8.6 features. The
8.6 band is much weaker than the 7.6/7.8 band, just as observed in `normal'
HII regions. In the Seyfert galaxy NGC\,1068, however, the 8.6 band is similar
in strength to the 7.7 band, as it is seen in the
ultracompact HII regions in M\,17 (Cesarsky 1996b) or IRAS\,18323-0242
(Roelfsema et al. 1996), where the UV radiation field is
extremely strong.

NGC\,1068, like many AGNs, shows an
additional component of warm dust in the 10$\mu$m region. Unified models for
Seyfert galaxies predict a dusty torus which would emit at these mid-infrared
wavelengths (e.g. Pier \& Krolik 1992). Hence, an alternative interpretation
of the weak emission features on both sides of the silicate absorption might
be self-absorbed silicate emission from the torus, i.e. the emissions we 
identified as PAH might simply be wings of a wide silicate emission maximum, 
the center of which is suppressed by absorption.
However, the observed double peaks at 7.7/8.6 and 11.05/11.25, as well as
the distinct rise in flux near 7.3$\mu$m are not reproduced by torus models
and show that there must be some
real, although weak, PAH emission on top of the continuum.
The weakness of the PAH emission can be understood in terms of dilution by the
hot dust continuum and destruction by the intense AGN radiation field.

\begin{figure*}
\setcounter{figure}{2}
 \resizebox{\hsize}{!}{\includegraphics{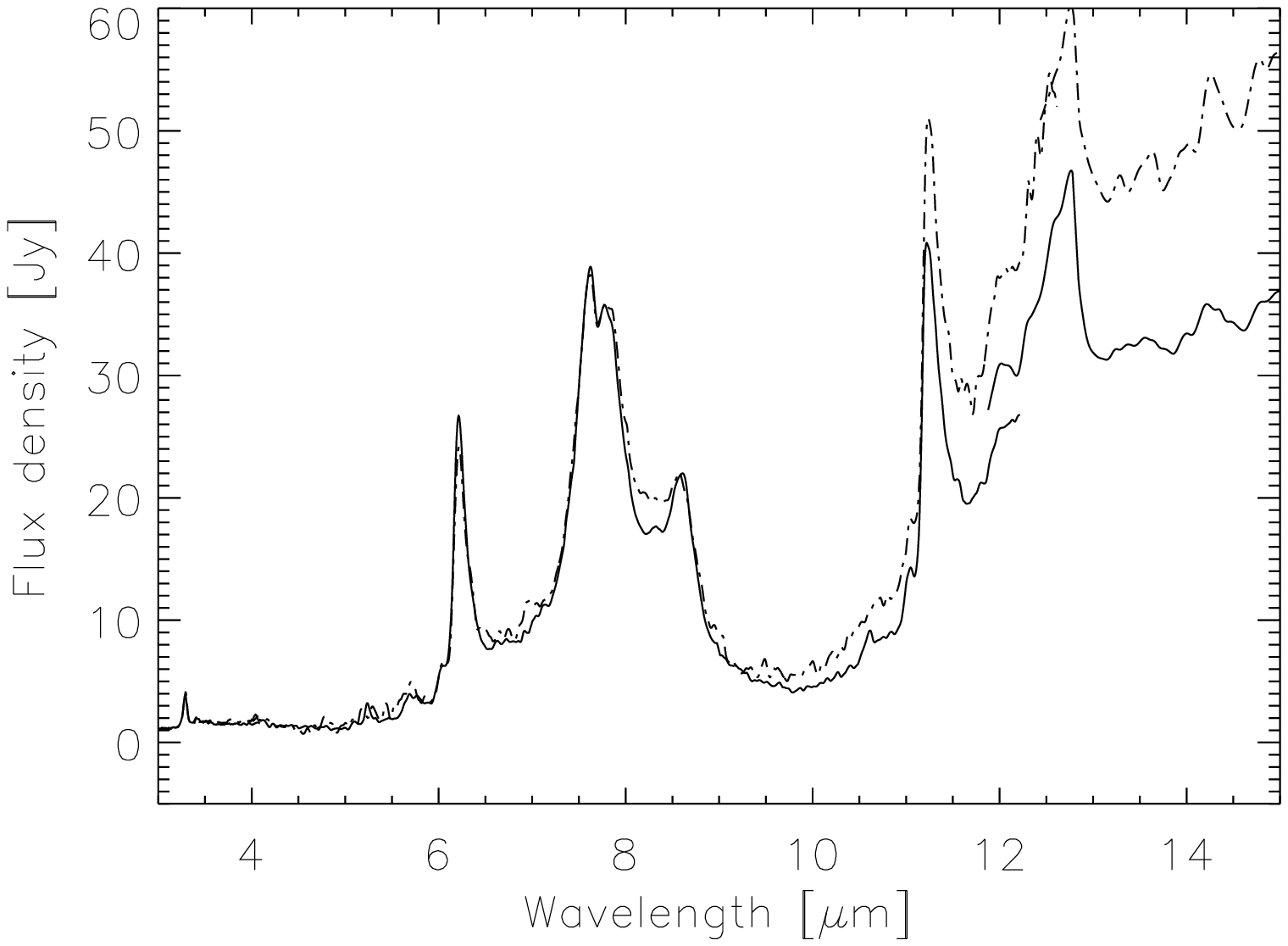}}
 \caption{M\,82 (full) versus NGC\,253 (dash-dotted). In both spectra narrow
 emission lines have been masked out. Both spectra have been smoothed and 
 corrected for zodiacal light and for their red shift. The spectrum of NGC\,253
 has been multiplied by 2.6 to normalize the 7.6$\mu$m PAH to the one in 
 M\,82.}
 \label{fig:m82vsngc253}
\end{figure*} 
\begin{figure*}
 \setcounter{figure}{3}
 \resizebox{\hsize}{!}{\includegraphics{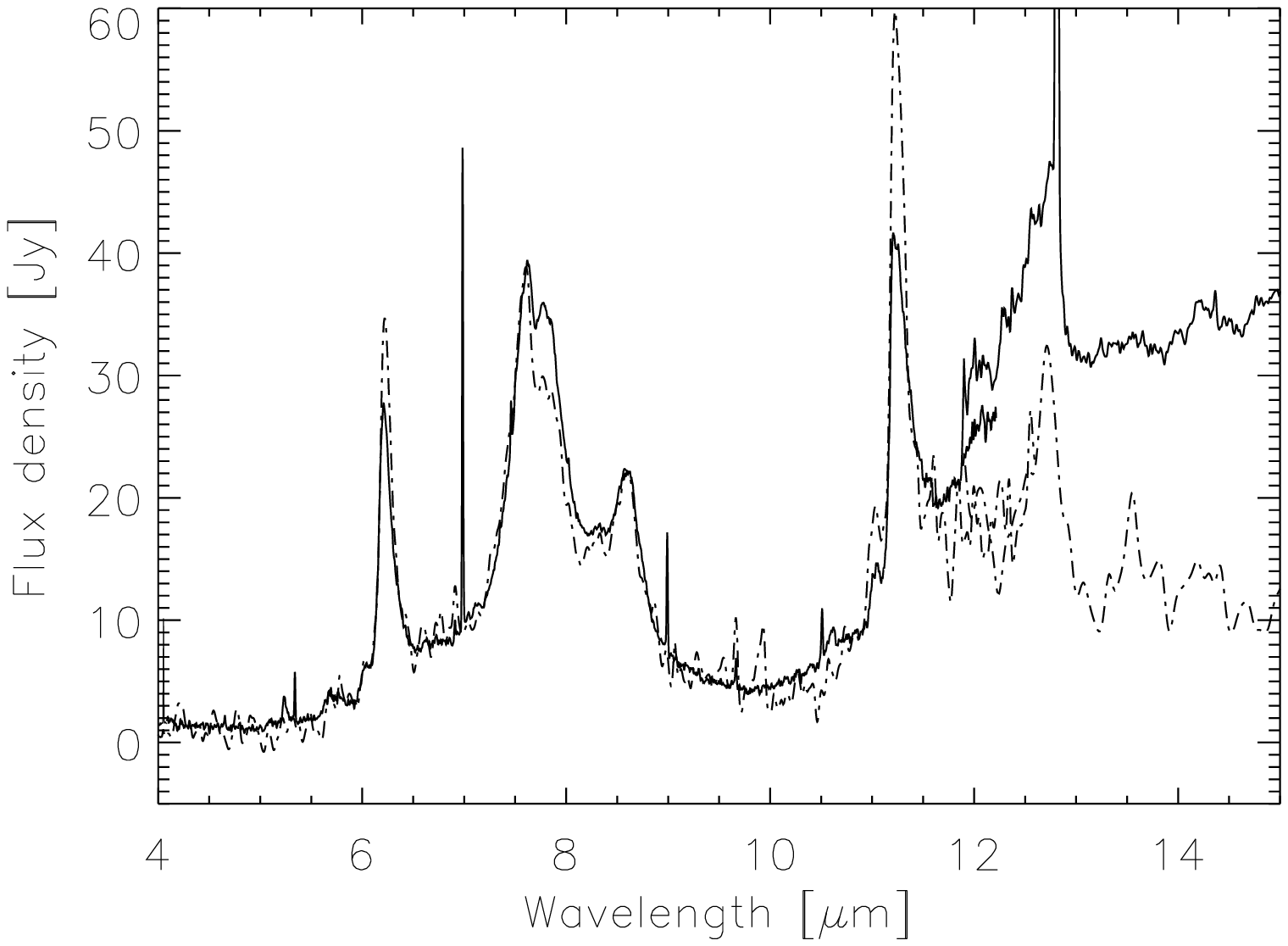}}
 \caption{M\,82 (full) versus the galactic reflection nebulae NGC\,7023
 (dash-dotted). Both spectra have been corrected
 for zodiacal light and for their red shift. The spectrum of NGC\,7023   
 has been smoothed and multiplied by 3.25 to normalize the 7.6$\mu$m PAH to 
 the one in M\,82.}
 \label{fig:m82vsngc7023}
\end{figure*} 
\begin{figure*}
 \setcounter{figure}{4}
 \resizebox{\hsize}{!}{\includegraphics{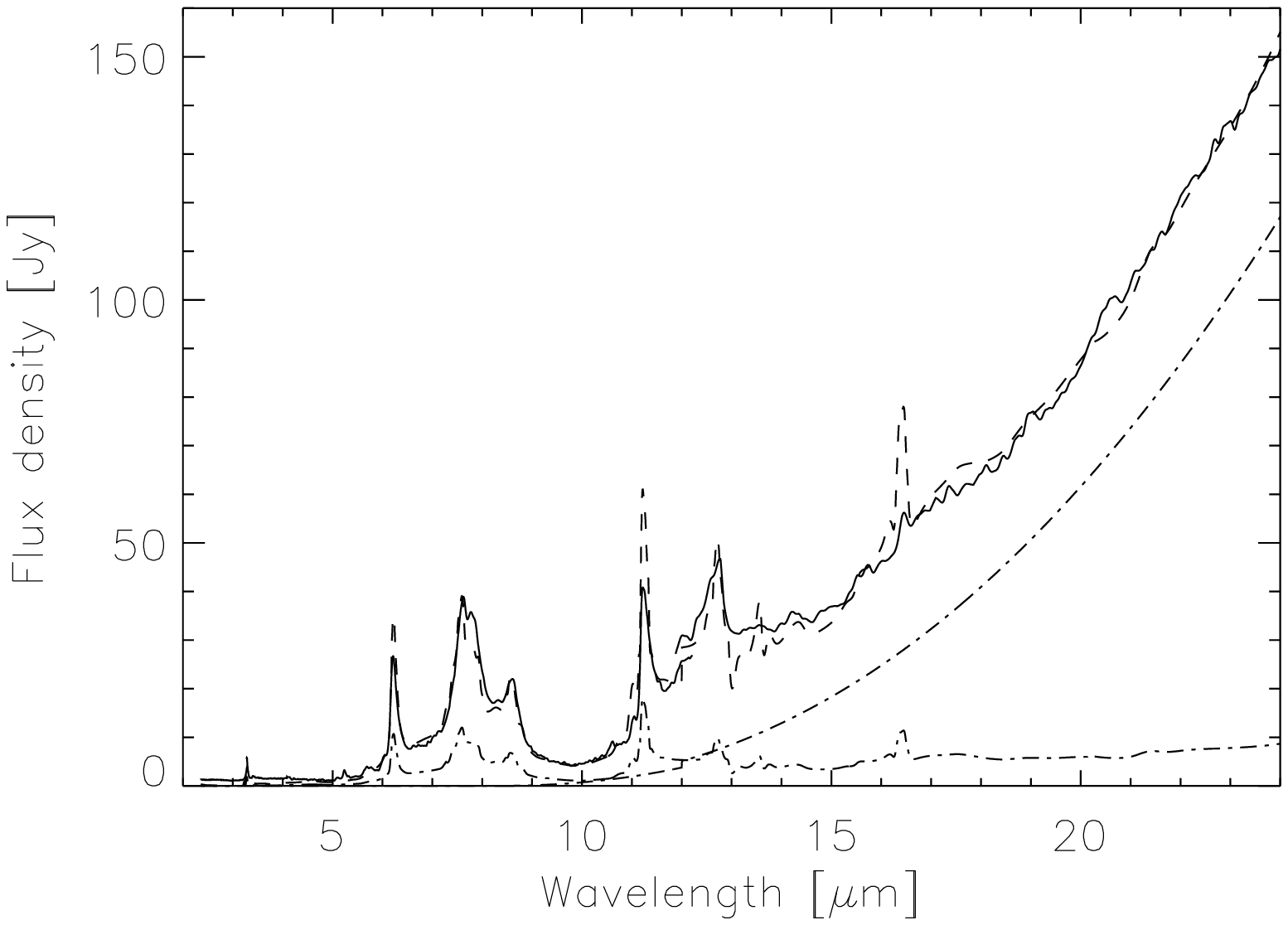}}
 \caption{M\,82 (full) versus NGC\,7023 x 3.25 + power law (dashed).
 The spectra of NGC\,7023 (dash-dotted) and M\,82 have been smoothed. 
 The power law component (dash-dotted) takes into account the aperture 
 change at 12$\mu$m (x 1.3).}
 \label{fig:m82sim}
\end{figure*} 

NGC\,1068 has a circumnuclear starburst region, and
some of the PAH emission might be picked up from this region by the large
SWS aperture. The match in shape with the (small-aperture)
ground-based data of Roche et al. (1984) in the 8-13$\mu$m range, and
comparisons to ground-based CO maps,
suggest that this effect is of minor importance here.

In the Circinus Galaxy the 7.7/8.6 ratio is somewhere in
between the two extrema M\,82 and NGC\,1068.
The feature/continuum ratio of the PAH features,
however, is much higher in Circinus than in NGC\,1068. We attribute this to
the fact that in the case of Circinus the large SWS aperture
indeed picked up parts of the well-known circumnuclear star forming region.

 

\section{Continuum placement and the depth of the silicate feature}
\label{s:continuum}

An important diagnostic in many extragalactic studies is the depth of the
silicate absorption feature at 9.7$\mu$m and the extinction derived from it.
However, the presence of strong PAH bands
on both sides of the 9.7$\mu$m
silicate feature makes it very difficult to estimate the continuum level
and the true depth of the silicate absorption.
In particular ground-based 8--13$\mu$m data suffer from this problem.
Because of their continuous wavelength coverage ISO-SWS01 spectra are
well suited to shed more light on this question.
In the following we will discuss this issue using the example of M\,82.

Firstly, evidence against a strong 9.7$\mu$m absorption comes 
from the absence of a strong 18$\mu$m silicate absorption (see Fig. 1).
According to Draine \& Lee (1984) the expected ratio 
$\tau_{sil}$(18$\mu$m) / $\tau_{sil}$(9.7$\mu$m) is 0.4.
Furthermore an analysis of hydrogen recombination lines in the ISO range
yields a moderate A$_V$(gas)$\approx$ 5 mag (for a uniform screen 
model - see Schreiber 1998). 

Next we come back to the comparison of the M\,82 spectrum to the spectrum of
NGC\,253 (Fig.
3). The two galaxies are similar
in A$_V$ and exhibit a very similar PAH spectrum. The only distinction appears
to be a stronger VSG continuum in NGC\,253. A strong
rise of the continuum in this range is typical for regions of intense
UV flux
(e.g. D\'esert et al. 1990, Vigroux et al. 1996) \footnote{The spectrum of 
NGC\,253 indicates lower ionization
(e.g. low [Ne III]/[[Ne II]), but due to the compactness of the starburst in
NGC\,253 the radiation field here is more intense than in M\,82.}.
The difference between both spectra
can be well fit by a power-law continuum that begins to rise at approximately
8--9$\mu$m.
 
Finally, we
compare our spectrum
of M\,82 to the SWS01 spectrum of the galactic reflection nebula NGC 7023.
Little extinction is expected in the line of sight to this nebula.
The continuum under the PAH bands
in NGC 7023 is very weak and starts to grow only beyond 20$\mu$m (Moutou et
al. 1998).
The flux density around 10$\mu$m is almost on the same level as the flux
density shortward of the 6$\mu$m PAH band and at 15--20$\mu$m.
It could be explained by an underlying PAH plateau, or by
the wings of the 7.7/8.6 and 11.3 PAH
features (which can be represented by Lorentz profiles - Boulanger et al. 1998,
Mattila et al. 1999).
We therefore assume that in NGC\,7023 this region consists
mainly of emission bands, with little or no continuum and silicate absorption.
In Fig.
4 we overplot the (smoothed) NGC\,7023 spectrum on that of M\,82. The NGC\,7023
spectrum has been multiplied by a factor 3.25 in order to normalize the
PAH emission feature at 7.6$\mu$m to the one in M\,82.
The 3.3 and 11.3$\mu$m bands
in M\,82 are weaker compared to NGC\,7023. This might be explained
e.g.
by the harder radiation field in M\,82 (Joblin et al. 1996, see 
Sect. \ref{s:PAH_var}).
Apart from this the two spectra are remarkably similar.
Only 
at higher wavelengths
a component of hot, small dust grains starts to add to the M\,82
continuum, as expected due to the much harder radiation field in M\,82.

In view of these arguments we constructed a toy model in order to reproduce
the observed
spectrum of M\,82. The model simply consists of the scaled spectrum of
NGC\,7023 plus a power-law continuum which starts at 8.5$\mu$m 
($f\propto(\lambda - 8.5)^{\alpha}$).
We use a power-law rather than a black body curve for sake
of simplicity. A power-law produces a very good fit to the continuum up to 
20 -- 25$\mu$m. A black-body curve would be more problematic, because
the dust may not have a single temperature, and might not be in thermal 
equilibrium.
M\,82 is an extended source for the SWS apertures. There is a flux jump
around 12$\mu$m by a factor of about 1.3, corresponding to a similar change
in aperture size. 
We hence multiplied the power law continuum by 1.3 longward of 12$\mu$m
\footnote{A more accurate correction for the change in aperture size would have
to assume a light distribution (as a function of wavelength) and a model of
the instrument beam profile. Such a correction tool is not yet available.}.
The free parameters - the scaling factors for the
NGC\,7023 spectrum and the power law, plus the power law index - were adjusted 
by hand; we did not pursue a formal fit.
Fig.
5 shows, that this simple model matches the observed
spectrum remarkably well. There is no need to invoke any kind of extinction.
Due to the uncertainties in the spectra, however, there is room for a moderate
extinction,
in accordance with the results from the recombination line studies (A$_V$ 
$\approx$ 5 mag).
A slightly different power-law, modified by a modest amount of extinction,
could fit the spectrum equally well.
However,
a strong overall extinction, as deduced from the ground based 8-13$\mu$m data
(A$_V$ = 15--60 mag, Gillett et al. 1975), is clearly incompatible
with the new ISO data.  This is an example of the potential danger of
overestimating the silicate absorption depth in baseline-limited data.

\section{The interpretation of low resolution spectra}
\label{s:lowres}
\begin{figure*}
\setcounter{figure}{5}
 \resizebox{\hsize}{!}{\includegraphics{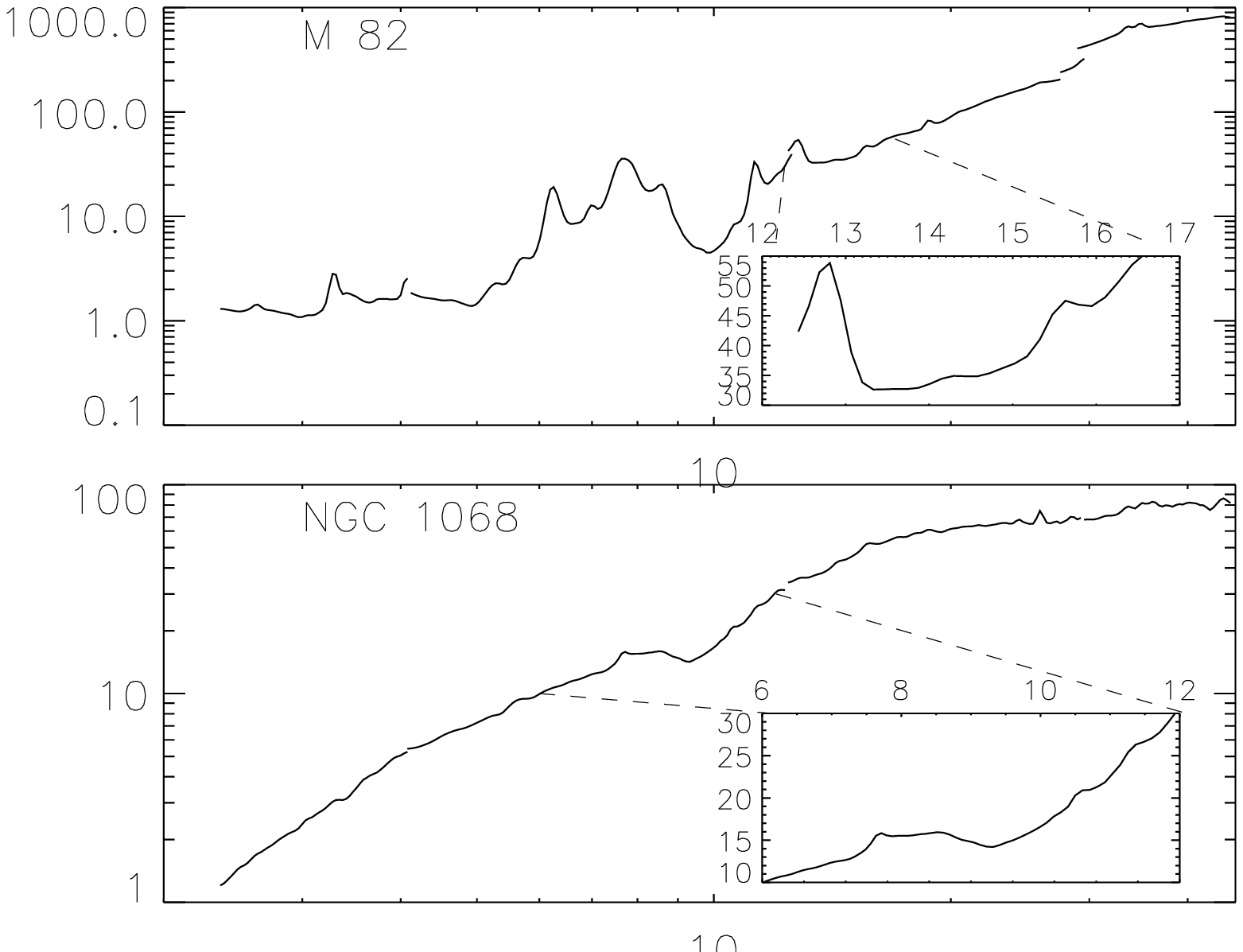}}
  \caption{The ISO-SWS spectra of M\,82 and NGC\,1068, 
  smoothed to a resolution of 50 
  to simulate the ISOCAM-CVF and SIRTF-IRS (low resolution mode) spectrometers.
  Wavelength in $\mu$m, flux density in
  Jansky.
  The feature at 14.3$\mu$m in M\,82 is NOT [Ne V] (see text).}
  \label{fig:m82lowres}
\end{figure*} 
Many ISO spectra of galactic and extragalactic objects have been
taken in low resolution mode (ISOPHOT-S, ISOCAM-CVF). Also, surveys
with future mid- and far-infrared space missions, e.g. of galaxies at higher
redshifts, will likely be
performed with a relatively low resolution. For example
the IRS spectrometer on board
SIRTF will have a resolution of approximately 50--100 
(plus a medium resolution mode of R=600) in a wavelength range similar to 
ISO-SWS.
In this low resolution mode SIRTF-IRS, being much more sensitive 
than ISO-SWS, will be a unique tool to detect emission features in spectra of 
faint high-z galaxies.
Low resolution spectra, however, suffer from possible identification
and interpretation problems caused by coincidences of
atomic/ionic lines and solid state features.
In our high resolution SWS01 galaxy spectra lines and features are well
separated, and we can use these spectra as templates to identify and highlight
the importance of possible confusion problems.
In Fig. 6
we have smoothed and rebinned the M\,82 and NGC\,1068 spectra
to a resolution of 50 to simulate e.g. an ISOCAM-CVF or a SIRTF-IRS spectrum.

\begin{table}
\caption[]{\label{tab:ne2_vs_pah} The flux ratio of PAH 12.7 / [Ne II] 12.8}
\begin{flushleft}
\begin{tabular}{ll}
\hline\noalign{\smallskip}
Object  & PAH 12.7/[Ne II] 12.8\\
\noalign{\smallskip}
\hline\noalign{\smallskip}
M\,82     & 0.96\\
NGC\,253  & 1.32\\
30\,Dor   & 0.00\\
Circinus  & 6.17\\
NGC\,1068 & 0..1$^a$\\
\noalign{\smallskip}
\hline
\end{tabular}
\end{flushleft}
\begin{list}{}{}
\item[$^{\rm a}$] exact value difficult to measure, due to the broad wing of 
                  the [Ne II] line and the relatively strong noise. 
\end{list}
\end{table}
In Table \ref{tab:inventory} we have indicated possible confusions with
nearby molecular, atomic, and ionic lines. We want to mention three lines in 
particular: 
[Ar II] at 6.99$\mu$m,  which might be 
confused with the underlying PAH emission (and the nearby H$_2$ S(5) line), 
[Ne II] at 12.8$\mu$m, which in 
fact has been confused in the past with the underlying 12.7$\mu$m PAH feature,
and [Ne V] at 14.3$\mu$m, which also has been
confused in the past with the nearby PAH emission.
To get an indication of the relative contributions of the 12.7$\mu$m 
PAH flux
and the [Ne II] line flux to the combined (line plus feature) flux in low 
resolution spectra
we have measured both fluxes in our high resolution spectra.
Table \ref{tab:ne2_vs_pah} summarizes the ratios of
PAH/[Ne II] in all 5 templates. The values
vary widely: in 30\,Dor
the flux is solely due to
[Ne II], whereas in Circinus [Ne II] contributes 
only about 15\% of the combined flux.  For the 7.0$\mu$m feature,
we find that the broad feature contributes 25\% of the combined flux of
feature, H$_2$, and [Ar II] in NGC\,253.

In the low resolution representation of
M\,82 in
Fig.
6 the shape of the 14.3$\mu$m 
PAH resembles very much the shape of an unresolved
line like the [Ne III] line
at 15.5$\mu$m and can be mistaken as [Ne V]. 
The high resolution spectrum of M\,82 (Fig. 2) clearly shows, that
there is no \mbox{[Ne V]} at 14.32$\mu$m but an 
emission feature plus a weak line of \mbox{[Cl II]} 14.37$\mu$m.
Of all the strong fine structure emission lines
only few lines remain unambiguously detectable in low resolution
spectra, like Br\,$\beta$, Br\,$\alpha$, [Ne III] 15$\mu$m,
\mbox{[S III]} 18.7, 33.5$\mu$m, and [Si II] 34.8$\mu$m  in M\,82, or
[O IV] 26$\mu$m and perhaps [Ne V] 24$\mu$m, and [S IV] 10.5$\mu$m in 
NGC\,1068.
Clearly, low resolution spectra are very well suited for PAH and continuum
measurements. However, flux measurements of narrow lines, and -- in some cases
-- even their identification, can be very difficult. For these purposes higher
resolutions, as for instance provided by the R=600 mode of the SIRTF 
spectrometer, are definitely needed.

\section{Conclusions}
\label{s:conclusions}

We have detected a large number of mid-infrared features in galaxy spectra, 
some of them previously unobserved, and discussed the dependence of the
dust features on ISM condition in galaxies.
The spectral features vary considerably from source to source in 
presence and relative strength. Emission features are largely absent in 
the intense radiation field close to an AGN, and weak in a
low metallicity, intensely star forming environment.
Differences in the absorption spectra point to different
physical properties of the obscuring regions in starburst and active galaxies.

The spectra presented here will be valuable template spectra for future
mid- and far-infrared space missions such as SIRTF, SOFIA or FIRST. 
They
provide important clues for the identification and interpretation of high
redshift, dusty galaxies.
The strongest PAH features can be used to provide redshift information in 
far-infrared photometric
galaxy surveys (Simpson \& Eisenhardt 1999, see also the example of
21396+3623, Rigopoulou et al. 1999). Furthermore, they affect
galaxy number counts.
For instance, Xu et al. (1998) have constructed semi-empirical galaxy SEDs
to model the considerable PAH effects on number counts and redshift distributions.
Finally, the continuum and the PAH features can be used to distinguish 
between starburst activity and active nuclei in high redshift galaxies, as
has been demonstrated for local infrared bright galaxies (Genzel et al. 1998,
Lutz et al. 1998a, Rigopoulou et al. 1999).

The advantage of the wide wavelength coverage of the SWS spectra has been
used to
illustrate the problem of the continuum definition and the true depth of the
silicate absorption.
We find that in our starburst templates the hot VSG dust continuum begins to
rise around 8 to 9$\mu$m,
and that it can be well fitted by a simple power-law up to 20...25$\mu$m.
Finally we have demonstrated possible line identification problems in low
resolution spectra.

The spectra presented here are available in electronic form from the authors.
We want to note again, that different parts of the spectra were
observed through different aperture sizes, which should be taken into account
for a detailed use as template spectra.

\begin{acknowledgements}
We wish to thank George Helou for very fruitful discussions, and Bernhard
Brandl for support with the SIRTF-IRS simulations.
      SWS and the ISO Spectometer Data Center at MPE are supported by
      DLR under grants 50 QI 8610 8 and 50 QI 9402 3. 
The ISO Spectral Analysis Package 
(ISAP) is a joint development by the LWS and SWS Instrument Teams and Data 
Centers. Contributing institutes are CESR,
IAS, IPAC, MPE, RAL and SRON.
\end{acknowledgements}


\end{document}